\documentclass[12pt,english,floatfix,nofootinbib,superscriptaddress,aps,prd,preprint]{revtex4}
\usepackage[utf8]{inputenc}       
\usepackage[english]{babel}       
\usepackage[utf8]{inputenc}
\usepackage{textcomp}
\usepackage{amsmath,amsthm,amssymb,amsfonts,mathrsfs,amsbsy} 
\usepackage{tensor}               
\usepackage{slashed}  
\usepackage{bbm}
\usepackage{esint}                
\usepackage{cancel}               
\usepackage{graphicx}             
\usepackage{natbib}
\usepackage{float}                
\usepackage{subfig}               
\usepackage[font=small,labelfont=bf]{caption} 
\usepackage{multirow}             
\usepackage{array}                
\usepackage{tikz}                 
\usetikzlibrary{quotes,angles,arrows,decorations.markings} 
\usepackage{makecell} 

\usepackage{lipsum}
\usepackage[a4paper, margin=1.6cm]{geometry}
\usepackage{xcolor}               
\usepackage{color}                
\usepackage{textcomp}             
\usepackage{units}                

\newcommand{\be}{\begin{equation}}
\newcommand{\ee}{\end{equation}}
\newcommand{\Be}{\begin{eqnarray}}
\newcommand{\Ee}{\end{eqnarray}}

\newcommand{\mincir}{\raise
-3.truept\hbox{\rlap{\hbox{$\sim$}}\raise4.truept\hbox{$<$}\ }}
\newcommand{\magcir}{\raise
-3.truept\hbox{\rlap{\hbox{$\sim$}}\raise4.truept\hbox{$>$}\ }}

\newcolumntype{Y}{>{\centering\arraybackslash}X}
\providecommand{\U}[1]

\usepackage[dvips]{epsfig}        
\usepackage[dvips]{graphicx}      
\usepackage{hyperref}             
\hypersetup{
    colorlinks=true,              
    breaklinks=true,              
    citecolor=blue,               
    linkcolor=[rgb]{0,0.5,0.9},   
    urlcolor=red,                 
    filecolor=green               
}

\newcommand{\ie}{\begin{equation}}
\newcommand{\fe}{\end{equation}}
\newcommand{\se}{\begin{eqnarray}}
\newcommand{\ff}{\end{eqnarray}}

\begin{document}

\title{Quantum features of a {non-commutative} Schwarzschild black hole}


\author{A. A. Ara\'{u}jo Filho}
\email{dilto@fisica.ufc.br}
\affiliation{Departamento de Física, Universidade Federal da Paraíba, Caixa Postal 5008, 58051--970, João Pessoa, Paraíba,  Brazil.}
\affiliation{Departamento de Física, Universidade Federal de Campina Grande Caixa Postal 10071, 58429-900 Campina Grande, Paraíba, Brazil.}
\affiliation{Center for Theoretical Physics, Khazar University, 41 Mehseti Street, Baku, AZ-1096, Azerbaijan.}

\author{I. P. Lobo}
\email{lobofisica@gmail.com}
\affiliation{Departamento de Física, Universidade Federal da Paraíba, Caixa Postal 5008, 58051--970, João Pessoa, Paraíba,  Brazil.}
\affiliation{Departamento de Física, Universidade Federal de Campina Grande Caixa Postal 10071, 58429-900 Campina Grande, Paraíba, Brazil.}


\author{P. H. M. Barros}
\email{phmbarros@ufpi.edu.br}
\affiliation{Departamento de F\'{i}sica, Universidade Federal do Piau\'{i} (UFPI), Campus Min. Petr\^{o}nio Portella, Teresina - PI, 64049-550, Brazil}
\affiliation{Departamento de Matemática e F\'{i}sica, Universidade  Estadual do Maranhão (UEMA), Caxias - MA, 65604-380, Brazil}

\author{Amilcar R. Queiroz}
\email{amilcarq@df.ufcg.edu.br}

\affiliation{Departamento de Física, Universidade Federal de Campina Grande Caixa Postal 10071, 58429-900 Campina Grande, Paraíba, Brazil.}

\date{\today}

\begin{abstract}

This work aims to present {some} quantum aspects of a {non--commutative (NC)} gauge gravity formulation of a Schwarzschild--like black hole constructed via the Moyal twist $\partial_t \wedge \partial_\theta$. Particle creation is estimated for bosonic and fermionic fields using the quantum tunneling method, with divergent integrals treated through the residue prescription. Since the surface gravity is well defined for this configuration, the corresponding emission rates and evaporation lifetimes are also computed. In addition, previously reported results in the literature on gauge gravity Schwarzschild black holes are revisited. Finally, we infer constraints on the {NC} parameter $\Theta$ from solar--system tests.

\end{abstract}


\maketitle

\tableofcontents


\section{Introduction }

Classical general relativity places no built-in bound on the precision with which distances can be resolved. By contrast, a wide class of quantum-gravity–motivated approaches point toward the emergence of a smallest measurable length, typically tied to the Planck scale. This expectation has led to the formulation of spacetime models in which the coordinates fail to commute, providing an effective way to encode such a cutoff at short distances \cite{szabo2003quantum,3,szabo2006symmetry}. Frameworks of this type have also become standard tools in the study of supersymmetric gauge field theories, where their structural and renormalization properties have been examined in detail \cite{ferrari2003finiteness}. When gravitational backgrounds are considered, NC corrections are commonly implemented through the Seiberg–Witten map, which deforms the underlying symmetry content of the geometry while preserving gauge consistency \cite{chamseddine2001deforming}.

Over the past years, NC constructions have been widely employed as an alternative route to explore black hole physics beyond the classical geometric picture \cite{mann2011cosmological,karimabadi2020non,zhao2024quasinormal,modesto2010charged,araujo2024effects,Anacleto:2019tdj,heidari2024exploring,araujo202as5properties,lopez2006towards,campos2022quasinormal}. Within this line of investigation, particular attention has been devoted to understanding how departures from ordinary spacetime commutativity reshape the dynamical evolution of black holes, with emphasis on the mechanisms governing particle emission and mass loss \cite{myung2007thermodynamics,AraujoFilho:2025rwr}. Alongside these dynamical aspects, the introduction of {NC} corrections has been shown to modify the thermodynamic description of black holes, motivating systematic analyses of quantities such as entropy, Hawking temperature, and heat capacity in a variety of scenarios \cite{sharif2011thermodynamics,nozari2007thermodynamics,lopez2006towards,banerjee2008NC,nozari2006reissner}.

A departure from classical spacetime geometry can be implemented by promoting the coordinates to non--commuting operators that satisfy the algebra $[x^\mu, x^\nu] = \mathbbm{i}\,\Theta^{\mu\nu}$, where the antisymmetric tensor $\Theta^{\mu\nu}$ encodes the scale and structure of the deformation. This modification replaces the conventional notion of a smooth manifold with an effective description in which short--distance behavior is intrinsically altered. Several formalisms have been proposed to accommodate such a deformation within gravitational theories. One particularly systematic construction is based on deforming the gauge symmetry underlying gravity: starting from a de Sitter–type group SO(4,1), the theory is consistently mapped onto the Poincaré group ISO(3,1) through the Seiberg–Witten prescription. Within this gauge--theoretic setting, Chaichian and collaborators~\cite{Chaichian:2007we} obtained a Schwarzschild--like solution whose metric explicitly reflects {NC} corrections.

Recent developments have reconsidered how black hole geometries are obtained in {NC} gauge theory. In particular, the analysis carried out in Ref.~\cite{Juric:2025kjl} reformulated the construction procedure and demonstrated that the method originally proposed in Ref.~\cite{Chaichian:2007we} was incomplete. The reassessment showed that the earlier formulation omitted a necessary contribution in the {NC} sector through the following term:
\ie
 - \frac{1}{16} \Theta^{\nu \rho} \Theta^{\lambda\tau} \Big[ \tilde{\omega}^{ac}_{\nu} \Tilde{\omega}^{cd}_{\lambda} \Big(  D_{\tau}R^{d5}_{\rho\mu} + \partial_{\tau}R^{d5}_{\rho\mu}   \Big)   \Big].
\nonumber
\fe
This omission propagated into the tetrad sector, producing nontrivial modifications of the vierbein fields and, as a direct consequence, altering the associated metric components.

Over the last year, a number of studies have addressed thermodynamics and particle creation in the framework of {NC} gauge theory \cite{Touati:2023gwv,Touati:2022zbm,Filho:2022zdh,Heidari:2023bww}. More recent analyses, however, have revisited the underlying construction and revealed the presence of inconsistencies in part of the existing literature \cite{Juric:2025kjl}. Along these lines, Refs.~\cite{AraujoFilho:2025iob,AraujoFilho:2025lxa} advanced the discussion by explicitly identifying and clarifying some of these shortcomings. Even so, the corrections proposed so far remain limited to specific Moyal twists, namely $\partial_r \wedge \partial_\theta$ and $\partial_t \wedge \partial_r$, for which the surface gravity is not well defined. As a result, some important quantities such as the thermodynamic functions, the evaporation lifetime, and the emission rate cannot be consistently determined in those configurations.

In order to resolve this limitation and address the open gap in the literature, the present work investigates quantum effects—specifically particle creation, evaporation, and emission rate—for a Schwarzschild--like black hole in {NC} gauge theory constructed with the Moyal twist $\partial_t \wedge \partial_\theta$. For this choice, in agreement with Ref.~\cite{Juric:2025kjl}, the surface gravity is well defined, which allows all these aspects to be treated consistently. In addition, to reinforce the physical robustness of the analysis, bounds on the {NC} parameter $\Theta$ are obtained from solar--system tests.


\section{The modified Schwarzschild black hole }

As it was pointed out above, Ref.~\cite{Juric:2025kjl} revisited and corrected the procedure previously proposed in Ref.~\cite{Chaichian:2007we} for constructing black hole solutions within {NC} gauge theory. As a consequence of this revision, all geometries obtained through that prescription are modified, as explicitly illustrated by the comparison performed for the Hayward spacetime~\cite{Heidari:2025iiv}. In particular, Ref.~\cite{Juric:2025kjl} shows that, for diagonal metrics, three distinct Schwarzschild--like solutions emerge depending on the choice of Moyal twist, namely $\partial_r \wedge \partial_\theta$, $\partial_t \wedge \partial_r$, and $\partial_t \wedge \partial_\theta$. Among these possibilities, only the $\partial_t \wedge \partial_\theta$ deformation leads to a well--defined surface gravity. Since the present analysis focuses on particle creation and evaporation, we adopt precisely this configuration to ensure the consistency of the thermodynamic quantities. The corresponding solution associated with the $(t,\theta)$ Moyal twist is therefore presented below \cite{Juric:2025kjl}:
\begin{equation}
    \begin{split}
        \mathrm{g}_{tt}(r,\Theta) &= -\left(1 - \frac{2M}{r}\right)
        + \frac{M^2(-2M + r)(-30M + 9r)}{48\,r^6} \Theta^{2}, \\
        \mathrm{g}_{rr}(r,\Theta) &= \left(1 - \frac{2M}{r}\right)^{-1}
        - \frac{M (11M^2 - 10Mr + 2r^2)}{4\,r^4(-2M + r)} \Theta^{2}, \\
        \mathrm{g}_{\theta \theta}(r,\Theta) &= r^2
        + \frac{M(11M - 4r)(2M - r)}{16\,r^3}\,\Theta^2, \\
        \mathrm{g}_{\varphi \varphi}(r,\Theta) &= r^2 \sin^2\theta
        + \frac{M(-2M + r)(M + r)\,\sin^2\theta}{4\,r^3} \Theta^{2}.
    \end{split}
    \label{eq1}
\end{equation}

\vspace{1em}
\noindent

The presence of the Moyal deformation $\partial_t \wedge \partial_\theta$ explicitly breaks the original spherical symmetry of the geometry. As a consequence, the resulting spacetime departs from the standard isotropic structure. The modified line element contains two distinguished radial loci: the origin at $r=0$, where the curvature diverges and a genuine singularity occurs, and the hypersurface $r=2M$, which continues to represent the event horizon and simultaneously defines a Killing horizon. Notably, the {NC} modification does not shift the horizon radius, which remains fixed at its ``classical value''. In addition, the corresponding surface gravity associated with this horizon is then expressed as \cite{Juric:2025kjl}
\begin{equation}
    \kappa^2 = \frac{M^2}{r_{h}^4}
    + \frac{\left(154 M^5 - 117 M^4 r_{h} + 20 M^3 r_{h}^2\right)}{16 r_{h}^9} \Theta^2
    + \mathcal{O}\left(\Theta^3\right).
    \label{sufaceee}
\end{equation}
Evaluating the geometry at $r_{h}=2M$ shows that the contribution proportional to $\Theta^{2}$ drops out identically, yielding a surface gravity $\kappa = 1/(4M)$. This result demonstrates that the {NC} parameter does not modify the surface gravity as well. Consequently, because both the horizon position and $\kappa$ remain unchanged under the $\Theta$ deformation, the thermodynamic quantities coincide with those of the standard Schwarzschild black hole. However, the particle creation process—both for bosonic and fermionic fields—as well as the evaporation lifetime and the emission rate, are affected by the {NC} parameter $\Theta$. The subsequent section is devoted to the investigation of these effects.


\section{Quantum aspects }


\subsection{Particle creation for bosons: {the tunneling approach} }

To incorporate backreaction effects associated with energy balance in the emission process, we adopt a semiclassical strategy inspired by earlier tunneling analyses \cite{011,vanzo2011tunnelling,parikh2004energy,Calmet:2023gbw}. The starting point consists in performing a coordinate transformation that casts the geometry into a Painlevé--Gullstrand–type representation, which is regular across the horizon \cite{parikh2004energy,Calmet:2023gbw}. In this framework, the line element assumes the form {
\ie
\mathrm{d}s^2 = - \mathrm{f}(r,\Theta)\,\mathrm{d}t^2 + 2 \mathrm{h}(r,\Theta) \,\mathrm{d}t \mathrm{d}r + \mathrm{d}r^2 + \mathrm{z}(r,\theta,\Theta)\,\mathrm{d}\theta^2 + \mathrm{y}(r,\theta,\Theta)\,\mathrm{d}\varphi^2,
\fe}
where the mixed component $\mathrm{h}(r,\Theta)$ is fixed by the requirement that the spatial slices remain flat, yielding \cite{Calmet:2023gbw} {
\ie
\mathrm{h}(r,\Theta) = \sqrt{\mathrm{f}(r,\Theta)\big(\mathrm{g}(r,\Theta)^{-1} - 1\big)}.
\fe}

Once the metric is expressed in this form, the quantum emission process follows from the nontrivial contribution to the imaginary sector of the particle’s action \cite{parikh2004energy,vanzo2011tunnelling}. The tunneling amplitude is therefore controlled by the classical action evaluated along a forbidden trajectory. For a massless excitation, this action is written as
$S = \int \mathrm{p}_\mu \, \mathrm{d}x^\mu.$

In evaluating the imaginary part of the action, the temporal sector does not generate any complex contribution. Indeed, the term $\mathrm{p}_t \mathrm{d}t = -\omega \mathrm{d}t$ remains strictly real along the trajectory and therefore plays no role in the emergence of an imaginary component. As a consequence, the complex part of the action originates solely from the integration along the radial direction, which provides the only nontrivial contribution to $\text{Im}\,S$
\ie
\text{Im}\,S =\text{Im}\,\int_{r_i}^{r_f} \,\mathrm{p}_r\,\mathrm{d}r=\text{Im}\,\int_{r_i}^{r_f}\int_{0}^{\mathrm{p}_r} \,\mathrm{d}\mathrm{p}_r'\,\mathrm{d}r.
\label{originalsss}
\fe

To incorporate the energy carried away during the quantum emission, the Hamiltonian is parametrized as $H = M - \omega'$, with $\omega'$ denoting the running energy of the escaping particle. 
{At fixed radial coordinate, Hamilton's equation,
$\dot{r}
=
\left.
\frac{\partial H}{\partial \mathrm{p}_r}
\right|_{r}$,
implies $\mathrm{d}\mathrm{p}_r'
=
\frac{\mathrm{d}H}{\dot{r}}$.
Moreover, the parametrization $H=M-\omega'$ gives
$\mathrm{d}H=-\,\mathrm{d}\omega'$. The running energy $\omega'$ is allowed to vary from $0$ to its final value $\omega$, which corresponds to the total energy carried away by the emitted particle.}
In this manner, the associated term entering the imaginary sector of the action can then be written in the following form:
\ie
\begin{split}
\text{Im}\, S
&=
\text{Im}\,\int_{r_i}^{r_f}\int_{M}^{M-\omega}
\,\frac{\mathrm{d}H}{\mathrm{d}r/\mathrm{d}t}\,\mathrm{d}r
{=
-\text{Im}\,\int_{0}^{\omega}\mathrm{d}\omega'
\int_{r_i}^{r_f}
\frac{\mathrm{d}r}{\mathrm{d}r/\mathrm{d}t}\,.}
\end{split}
\fe

{For radial null trajectories, characterized by $\mathrm{d}s^{2}=\mathrm{d}\theta=\mathrm{d}\phi=0$, the Painlev\'e--Gullstrand--type line element yields the ingoing and outgoing solutions
\[
\frac{\mathrm{d}r}{\mathrm{d}t}
=
-\mathrm{h}(r,\Theta)
\pm
\sqrt{\mathrm{f}(r,\Theta)+\mathrm{h}(r,\Theta)^{2}}.
\]
Since the tunneling probability for emission is determined by the outgoing trajectory, we select the positive branch, which explicitly reads}
\ie
\begin{split}
\frac{\mathrm{d}r}{\mathrm{d}t} =&  -\mathrm{h}(r,\Theta)+\sqrt{\mathrm{f}(r,\Theta)+\mathrm{h}(r,\Theta)^2} \\
& =-\frac{\sqrt{-M \left(\Theta ^2 M^2 (10 M-3 r)+16 r^5\right) \left(11 \Theta ^2 M^2+2 \Theta ^2 r (r-5 M)-8 r^4\right)}}{8 r^5} \\
& +\frac{\sqrt{\left(\Theta ^2 M^2 (10 M-3 r)+16 r^5\right) \left(\Theta ^2 M \left(-11 M^2+10 M r-2 r^2\right)+4 r^5\right)}}{8 r^5}.
\end{split}
\fe

Here, the backreaction induced by the emitted quantum is incorporated by promoting the mass parameter to a dynamical quantity. This is achieved through the shift $M \to (M - \omega')$, which alters the background geometry accordingly and
{leads, after expanding the outgoing radial velocity up to second order in $\Theta$, to the following expression for the imaginary part of the action:}
\ie
\begin{split}
\label{ims}
& \text{Im}\,S \approx -\text{Im}\,\int_{0}^{\omega}\mathrm{d}\omega' \int_{r_i}^{r_f} \frac{\mathrm{d}r} { 1-\sqrt{2}\sqrt{\dfrac{M-\omega'}{r}} + \Theta^{2}\Xi(r,M,\omega') }\,.
\end{split}
\fe
where {terms beyond second order in $\Theta$ have been neglected. For convenience, the auxiliary function $\Xi$ is defined by}
\ie
\begin{split}
{\Xi (r,M,\omega')} = & \frac{-34 (M-\omega')^3+37 (M-\omega')^2 r-8 (M-\omega') r^2}{32 r^5}\\
& +\frac{\sqrt{(M-\omega')} \left(-10 (M-\omega')^3+25 (M-\omega')^2 r-20 (M-\omega') r^2+4 r^3\right)}{16 \sqrt{2} r^{11/2}}
\label{sigmmma}
\end{split}
\fe

{
For each fixed value of $\omega'$, we define $\mu=M-\omega'>0$, so that the horizon is located at $r=2\mu$. Before expanding the denominator of Eq.~(\ref{ims}), let us explicitly examine the behavior of the auxiliary function $\Xi$ in this limit. Direct substitution into Eq.~(\ref{sigmmma}) gives $\Xi(2\mu,M,\omega')=1/(128\mu^{2})-1/(128\mu^{2})=0$. Moreover, taking the radial derivative at fixed $M$ and $\omega'$, we obtain $\left.\partial_r\Xi(r,M,\omega')\right|_{r=2\mu}=-15/(1024\mu^{3})+15/(1024\mu^{3})=0$. These cancellations show that both the auxiliary function and its first derivative vanish at the horizon. Accordingly, its Taylor expansion reads $\Xi(r,M,\omega')=-3(r-2\mu)^{2}/(1024\mu^{4})+\mathcal{O}\!\left((r-2\mu)^{3}\right)$. The contribution proportional to $\Theta^{2}$ begins, in this manner, at quadratic order in $r-2\mu$, leaving the linear coefficient of the denominator unchanged. In addition, the corresponding correction to the radial integrand, $-\Theta^{2}\Xi(r,M,\omega')/(1-\sqrt{2\mu/r})^{2}$, approaches the finite value $3\Theta^{2}/(64\mu^{2})$ as $r\to2\mu$ and contributes no pole residue, as we could naturally expect. In this regard, through second order in $\Theta$, the linearized denominator takes the following form:}
\ie 
1-\sqrt{\frac{2\mu}{r}}
+\Theta^{2}\Xi(r,M,\omega') \simeq \left( \frac{1}{4\mu}  \right) (r-2\mu).
\fe
Accordingly, the residue of the radial integrand at the instantaneous horizon is
\ie
\operatorname*{Res}_{r=2\mu} \left[ \frac{1}{ 1-\sqrt{\dfrac{2\mu}{r}} +\Theta^{2}\Xi(r,M,\omega') } \right] \simeq 4\mu.
\fe
{Applying the Feynman prescription and deforming the contour around the pole in the counterclockwise direction, the radial integration gives}
\ie
\begin{split}
{ \text{Im}\,S} & {=
\int_{0}^{\omega} \left[ 4\pi(M-\omega') \right] \mathrm{d}\omega'= 2\pi\omega(2M-\omega).}
\end{split}
\fe

Following this procedure, the probability for the emission of a \textit{Hawking} quantum—now incorporating the {NC} modifications—can be written in the form:
\ie
\Gamma (\omega,\Theta) \sim e^{-2 \, \text{Im}\, S} = { e^{- 4 \pi  \omega  (2 M-\omega )  } }.
\fe
Moreover, the occupation number (particle number density) can be written directly in terms of the tunneling probability as follows:
\ie
n(\omega,\Theta) = {\frac{\Gamma(\omega,\Theta)}{1 - \Gamma(\omega,\Theta)} = \frac{1}{\exp \left\{4 \pi  \omega  (2 M-\omega )\right\}-1}}.
\fe
{In other words, the bosonic particle density receives no correction from the NC parameter. This result is expected from Eq.~(\ref{sufaceee}), which shows that the surface gravity is unaffected by $\Theta$ and, consequently, that no corresponding correction to the Hawking radiation should arise. To independently verify this conclusion, in the next subsection we investigate particle creation using the near--horizon method within a WKB--like approximation.   }


{\subsection{Particle creation for bosons: {near horizon method} }}

{

To analyze bosonic particle creation through the same near-horizon procedure employed for fermions, we consider a scalar field of mass $m$ propagating on the background described by Eq.~(\ref{eq1}). Writing
$g_{tt}(r,\Theta)=-A(r,\Theta)$, $g_{rr}(r,\Theta)=B(r,\Theta)$, $g_{\theta\theta}(r,\Theta)=C(r,\Theta)$ and $ g_{\phi\phi} (r,\theta,\Theta)=D(r,\theta,\Theta)$,
the scalar field satisfies the curved spacetime Klein--Gordon equation
\begin{equation}
\frac{1}{\sqrt{-g}}\, \partial_{\mu}\left(
\sqrt{-g}\,g^{\mu\nu}\partial_{\nu}\Phi\right)
-\frac{m^{2}}{\hbar^{2}}\Phi=0.
\label{eq:KG-boson}
\end{equation}

In the semiclassical approximation, we adopt the WKB ansatz \cite{Maluf:2018lyu}
\begin{equation}
\Phi(x,\Theta) = \mathcal{A}(x,\Theta)
\exp\left[ \frac{i}{\hbar}I(x,\Theta) \right],
\label{eq:WKB-boson}
\end{equation}
where $\mathcal{A}(x,\Theta)$ is a slowly varying amplitude and $I(x,\Theta)$ is the classical action. Substituting Eq.~\eqref{eq:WKB-boson} into Eq.~\eqref{eq:KG-boson} and retaining only the leading terms in $\hbar$ gives the Hamilton--Jacobi equation
\begin{equation}
g^{\mu\nu}\partial_{\mu}I\,\partial_{\nu}I+m^{2}=0.
\label{eq:HJ-covariant-boson}
\end{equation}
For the diagonal metric under consideration, this equation becomes
\begin{equation}
-\frac{1}{A(r,\Theta)} \left(\partial_{t}I\right)^{2} + \frac{1}{B(r,\Theta)} \left(\partial_{r}I\right)^{2} + \frac{1}{C(r,\Theta)} \left(\partial_{\theta}I\right)^{2} + \frac{1}{D(r,\theta,\Theta)} \left(\partial_{\phi}I\right)^{2} + m^{2} =0.
\label{eq:HJ-expanded-boson}
\end{equation}

We separate the action in the form
\begin{equation}
I(x,\Theta) = -\omega t+W(r,\Theta)+J(\theta,\phi),
\label{eq:action-boson}
\end{equation}
where $\omega$ is the conserved energy of the emitted particle. The radial part then satisfies
\begin{equation}
W'_{\pm}(r,\Theta) = \pm \sqrt{ B(r,\Theta) \left[ \frac{\omega^{2}}{A(r,\Theta)} -m^{2} -\frac{\left(\partial_{\theta}J\right)^{2}}{C(r,\Theta)} -\frac{\left(\partial_{\phi}J\right)^{2}} {D(r,\theta,\Theta)} \right] }.
\label{eq:radial-general-boson}
\end{equation}
The signs $+$ and $-$ describe outgoing and ingoing trajectories, respectively. Near the event horizon, the term proportional to $\omega^{2}/A(r,\Theta)$ dominates, whereas the mass and angular terms remain finite. Therefore,
\begin{equation}
W'_{\pm}(r,\Theta) \simeq \pm\omega \sqrt{\frac{B(r,\Theta)}{A(r,\Theta)}} = \pm \frac{\omega} {\sqrt{A(r,\Theta)/B(r,\Theta)}}.
\label{eq:radial-near-horizon-boson}
\end{equation}

Close to $r=r_{h}$, the relevant metric functions can be expanded as
\begin{equation}
A(r,\Theta) \simeq A'(r_{h},\Theta)(r-r_{h}),
\end{equation}
and
\begin{equation}
\frac{1}{B(r,\Theta)} \simeq \left. \frac{\mathrm{d}}{\mathrm{d}r} \left( \frac{1}{B(r,\Theta)} \right) \right|_{r=r_{h}} (r-r_{h}).
\label{eq:metric-expansion-boson}
\end{equation}
{Then},
\begin{equation}
\sqrt{\frac{A(r,\Theta)}{B(r,\Theta)}} \simeq 2\kappa(r_{h},\Theta)(r-r_{h}),
\label{eq:AB-near-horizon-boson}
\end{equation}
where
\begin{equation}
\kappa(r_{h},\Theta) = \frac{1}{2} \sqrt{ A'(r_{h},\Theta) \left. \frac{\mathrm{d}}{\mathrm{d}r} \left( \frac{1}{B(r,\Theta)} \right) \right|_{r=r_{h}} }.
\label{eq:kappa-definition-boson}
\end{equation}

For the metric in Eq.~(1), the event horizon is located at $r_{h}=2M$. The derivatives entering Eq.~\eqref{eq:kappa-definition-boson} are
\begin{equation}
A'(r_{h},\Theta) = \frac{1}{2M} + \frac{\Theta^{2}}{256M^{3}},
\label{eq:A-derivative-boson}
\end{equation}
and
\begin{equation}
\left. \frac{\mathrm{d}}{\mathrm{d}r} \left( \frac{1}{B(r,\Theta)} \right) \right|_{r=r_{h}} = \frac{1}{2M} - \frac{\Theta^{2}}{256M^{3}}.
\label{eq:B-derivative-boson}
\end{equation}
The corrections of second order in $\Theta$ cancel in the product of Eqs.~\eqref{eq:A-derivative-boson} and \eqref{eq:B-derivative-boson}. Therefore,
\begin{equation}
\kappa(r_{h},\Theta) = \frac{1}{4M} + \mathcal{O}\left(\Theta^{4}\right).
\label{eq:kappa-result-boson}
\end{equation}

{
The cancellation in Eq.~(\ref{eq:kappa-result-boson}) deserves further clarification. Writing $F(r,\Theta)=1/B(r,\Theta)$, Eqs.~(\ref{eq:A-derivative-boson}) and (\ref{eq:B-derivative-boson}) show that the second order contributions to the horizon derivatives are equal in magnitude and opposite in sign, namely $\delta A'(2M) = \Theta^{2}/(256M^{3})$ and $\delta F'(2M) = -\Theta^{2}/(256M^{3})$. Since their zeroth order values coincide, the correction to their product vanishes at this order. This establishes the cancellation for the $(t,\theta)$ twist considered here. However, the present calculation does not establish a general identity of the Seiberg--Witten map that would enforce the same result at higher orders.

Although the horizon area also retains its Schwarzschild value, $\mathcal{A}_{h}=16\pi M^{2}$, through second order in $\Theta$, this fact alone does not explain the cancellation in the surface gravity. In particular, it is worthy to be mentioned that a Wald entropy argument would require evaluating the entropy associated with the deformed gravitational action and establishing the corresponding first law. Such a derivation is not supplied by the metric alone, and we do not assume that the unchanged horizon area guarantees an unchanged entropy in the deformed theory.

A calculation through fourth order would provide a further test of whether this cancellation persists. However, the deformed geometry adopted in the present work is taken from Ref.~\cite{Juric:2025kjl}, where the Seiberg--Witten construction is explicitly developed through second order in $\Theta$. Our results are fundamentally restricted to the perturbative order at which this geometry has been obtained. Determining the contribution at order $\Theta^{4}$ would require extending that construction and deriving the corresponding metric corrections, including any change in the horizon position. Such a contribution cannot be determined simply by retaining only products of the second order terms already available. Accordingly, Eq.~(\ref{eq:kappa-result-boson}) establishes the absence of a correction at order $\Theta^{2}$, while whether this cancellation persists at fourth order remains an open question.
}

In this manner, consistently with the surface gravity calculation presented in Sec.~II, the NC deformation does not modify the Hawking temperature at second order in $\Theta$:
\begin{equation}
T_{H} = \frac{\kappa}{2\pi} = \frac{1}{8\pi M} + \mathcal{O}\left(\Theta^{4}\right).
\label{eq:temperature-boson}
\end{equation}

Using Eqs.~\eqref{eq:radial-near-horizon-boson} and \eqref{eq:AB-near-horizon-boson}, the radial actions can be written as
\begin{equation}
W_{\pm}(r,\Theta) \simeq \pm \frac{\omega}{2\kappa(r_{h},\Theta)} \int \frac{\mathrm{d}r}{r-r_{h}}.
\label{eq:radial-integral-boson}
\end{equation}
Applying the Feynman prescription to regulate the pole at the horizon gives
\begin{equation}
2\,\mathrm{Im} \left[ W_{\mathrm{out}}(r,\Theta) - W_{\mathrm{in}}(r,\Theta) \right] = \frac{2\pi\omega}{\kappa(r_{h},\Theta)}.
\label{eq:imaginary-action-boson}
\end{equation}
The corresponding tunneling probability is therefore
\begin{equation}
\Gamma_{\Phi}(\omega,\Theta) \sim \exp\left[ -\frac{2\pi\omega}{\kappa(r_{h},\Theta)} \right] = \exp\left[ -8\pi M\omega + \mathcal{O}\left(\Theta^{4}\right) \right].
\label{eq:tunneling-probability-boson}
\end{equation}

For bosonic particles, the associated occupation number follows the Bose--Einstein distribution:
\begin{equation}
n_{\Phi}(\omega,\Theta) = \frac{\Gamma_{\Phi}(\omega,\Theta)} {1-\Gamma_{\Phi}(\omega,\Theta)} = \frac{1} { \exp\left[ 8\pi M\omega + \mathcal{O}\left(\Theta^{4}\right) \right] -1 }.
\label{eq:occupation-number-boson}
\end{equation}

}


\subsection{Particle production for fermions: {near horizon approximation} }

The semiclassical description of fermion emission from a black hole horizon is developed by treating the spinor field as a probe propagating on a fixed curved background. Instead of introducing horizon–regular coordinates at the outset—an approach commonly adopted in earlier tunneling analyses based on Painlevé–Gullstrand or Kruskal–Szekeres constructions—the formulation here relies directly on the covariant field equations in the original coordinate system \cite{o69}.

The behavior of a massive spin–$\tfrac{1}{2}$ field in this geometry follows from the curved–spacetime Dirac equation, {
\ie
\left(\bar{\gamma}^{\mu}\bar{\nabla}_{\mu}+m\right)\bar{\Psi}^{\pm}_{\uparrow\downarrow}(x,\Theta)=0,
\fe}
where the {NC} parameter $\Theta$ enters implicitly through the metric functions. The spinor covariant derivative is defined as
$\bar{\nabla}_{\mu}=\partial_{\mu}+\frac{\mathbbm{i}}{2}{\Gamma^{\alpha}}_{\mu}{}^{\beta},\bar{\Sigma}_{\alpha\beta}$,
with the Lorentz generators given by
$\bar{\Sigma}_{\alpha\beta}=\frac{\mathbbm{i}}{4},[\bar{\gamma}_{\alpha},\bar{\gamma}_{\beta}]$.

The spacetime coordinates are chosen as $x=(t,r,\theta,\varphi)$. The matrices $\bar{\gamma}^{\mu}$ provide a curved–space representation of the Clifford algebra and satisfy the anticommutation relation
$\{\bar{\gamma}_{\alpha},\bar{\gamma}_{\beta}\}=2\,g_{\alpha\beta}\,\mathbbm{1}$,
where $\mathbbm{1}$ denotes the $4\times4$ identity matrix.
For later convenience, the metric coefficients are rewritten in the form
$g_{tt}(r,\Theta)\equiv -A(r,\Theta)$,
$g_{rr}(r,\Theta)\equiv B(r,\Theta)$,
$g_{\theta\theta}(r,\Theta)\equiv C(r,\Theta)$,
and
$g_{\varphi\varphi}(r,{\theta,}\Theta)\equiv D(r,{\theta,}\Theta)$.

This parametrization isolates the radial structure responsible for horizon effects and prepares the ground for the semiclassical evaluation of fermionic tunneling probabilities. A suitable explicit representation of the $\bar{\gamma}$ matrices is then introduced to carry out the calculation
\begin{eqnarray}
 \Bar{\gamma}^{t}(r,\Theta) &=&\frac{\mathbbm{i}}{\sqrt{{A{(r,\Theta)}}}}\left( \begin{array}{cc}
\bf{1}& \bf{ 0} \\ 
\bf{ 0} & -\bf{ 1}%
\end{array}%
\right), \;\;
\Bar{\gamma}^{r}(r,\Theta) =\sqrt{\frac{1}{B{(r,\Theta)}{}}}\left( 
\begin{array}{cc}
\bf{0} &  \sigma_{3} \\ 
 \sigma_{3} & \bf{0}%
\end{array}%
\right), \\
\Bar{\gamma}^{\theta }(r,\Theta) &=&\frac{1}{\sqrt{{C{(r,\Theta)}}}}\left( 
\begin{array}{cc}
\bf{0} &  \sigma_{1} \\ 
 \sigma_{1} & \bf{0}%
\end{array}%
\right), \;\;
\Bar{\gamma}^{\varphi }(r,\Theta) =\frac{1}{{\sqrt{D{(r,{\theta,}\Theta)}} }}\left( 
\begin{array}{cc}
\bf{0} &  \sigma_{2} \\ 
 \sigma_{2} & \bf{0}%
\end{array}%
\right).
\end{eqnarray}%
In the present formulation, the matrices $\sigma$ are introduced as the generators of the two–dimensional spinor sector. Their multiplication law follows directly from their defining properties and can be compactly expressed as
$\sigma_i \sigma_j = \delta_{ij}\,\mathbbm{1} + \mathbbm{i}\,\varepsilon_{ijk}\,\sigma_k$,
with the indices $i,j,k$ running over $1,2,3$.
Beyond this algebraic structure, the chirality operator acquires a modified form due to the influence of non--commutativity. In this context, the matrix $\gamma^5(r,\Theta)$ is no longer treated as a purely constant object but instead reflects the underlying geometric deformation. Its explicit realization is therefore identified with the following matrix structure:
\begin{equation}
\gamma^{5}(r,\Theta) = \mathbbm{i} \gamma^{t}(r,\Theta)\gamma^{r}(r,\Theta)\gamma^{\theta }(r,\Theta)\gamma^{\varphi}(r,\Theta) = \mathbbm{i}\sqrt{\frac{1}{{A{(r,\Theta)} \, B{(r,\Theta)} \, C{(r,\Theta)} \, D{(r,{\theta,}\Theta)} }}}\left( 
\begin{array}{cc}
\bf{ 0} & - \bf{ 1} \\ 
\bf{ 1} & \bf{ 0}%
\end{array}%
\right)\:.
\end{equation}

A fermionic configuration with spin aligned along the outward radial direction is described by adopting the following spinor ansatz \cite{vanzo2011tunnelling}:
\begin{equation}
\psi^{(+)}_{\uparrow}(x,\Theta) = \left( \begin{array}{c}
H^{(+)}_{\uparrow}(x,\Theta) \\ 
0 \\ 
Y^{(+)}_{\uparrow}(x,\Theta) \\ 
0
\end{array}
\right) \exp \left[ \mathbbm{i} \, \psi^{(+)}_{\uparrow}(x,\Theta)\right]\;.
\label{spinupbh} 
\end{equation}

The discussion is restricted to the sector in which the fermionic spin is aligned with the outward radial direction, labeled by the $(+)$ polarization. The opposite orientation, denoted by $(-)$, follows from the same reasoning and therefore is not treated separately. Upon substituting the corresponding spinor ansatz, given in Eq.~(\ref{spinupbh}), into the curved–space Dirac equation, a coupled system of relations emerges. This step follows the semiclassical tunneling prescription developed in Ref.~\cite{vanzo2011tunnelling}, although the present formulation is adapted to the geometry under consideration.

In carrying out the calculation, only the dominant terms in the $\hbar$ expansion are kept, while higher--order quantum corrections are discarded. To separate the dynamical variables, the classical action is assumed to take the form
$\psi^{(+)}_{\uparrow}=-\omega\,t+\chi(r,\Theta)+L(\theta,\varphi)$,
which allows the temporal, radial, and angular contributions to be treated independently, leading to the relations discussed below
\cite{vanzo2011tunnelling} 
\begin{align}
&+\frac{\mathbbm{i}\, \omega H^{(+)}_{\uparrow}(x,\Theta)}{\sqrt{A{(r,\Theta)}}} 
-\sqrt{\frac{1}{B{(r,\Theta)}}} Y^{(+)}_{\uparrow}(x,\Theta)\,\chi^{\prime}(r,\Theta)
+m H^{(+)}_{\uparrow}(x,\Theta) = 0, \label{eq11} \\[10pt]
& - Y^{(+)}_{\uparrow}(x,\Theta) \left(  \frac{\partial_{\theta}L(\theta,\varphi)}{\sqrt{C{(r,\Theta)}}}
+\frac{\mathbbm{i}\,\partial_{\varphi} L(\theta,\varphi)}{\sqrt{D{(r,{\theta,}\Theta)}}} \right) = 0, \label{eq21} \\[10pt]
&    -\frac{\mathbbm{i} \, \omega Y^{(+)}_{\uparrow}(x,\Theta)}{\sqrt{A{(r,\Theta)}}}
-\sqrt{\frac{1}{B{(r,\Theta)}}}H^{(+)}_{\uparrow}(x,\Theta)\chi^{\prime}(r,\Theta)
+ m  Y^{(+)}_{\uparrow}(x,\Theta) = 0, \label{eq31} \\[10pt]
& - H^{(+)}_{\uparrow}(x,\Theta) \left(  \frac{\partial_{\theta}L(\theta,\varphi)}{\sqrt{C{(r,\Theta)}}}
+\frac{\mathbbm{i}\,\partial_{\varphi} L(\theta,\varphi)}{\sqrt{D{(r,{\theta,}\Theta)}}} \right) = 0. \label{eq41}
\end{align}

The explicit functional forms of $H^{(+)}_{\uparrow}(x,\Theta)$ and $Y^{(+)}_{\uparrow}(x,\Theta)$ do not play a decisive role in what follows. What matters instead is that the simultaneous validity of Eqs.~(\ref{eq21}) and (\ref{eq41}) imposes a nontrivial restriction on the angular sector, namely
$\frac{\partial_{\theta}L(\theta,\varphi)}{\sqrt{C^{(r,\Theta)}}}
+\frac{\mathbbm{i},\partial_{\varphi}L(\theta,\varphi)}{\sqrt{D^{(r,{\theta,}\Theta)}}}=0.$
This condition forces the angular contribution $L(\theta,\varphi)$ to be complex. The same requirement arises for both emission and absorption channels, implying that the angular dependence contributes identically to ingoing and outgoing solutions. Consequently, such terms drop out when forming the ratio of tunneling probabilities, and they can be omitted from the remainder of the computation. This cancellation mechanism is consistent with earlier semiclassical analyses of fermionic tunneling \cite{vanzo2011tunnelling}.

Specializing now to the massless limit, $m=0$, the remaining radial equations decouple and admit two independent branches. These solutions are characterized by the relations
$H^{(+)}_{\uparrow}(x,\Theta)=-\mathbbm{i}\,Y^{(+)}_{\uparrow}(x,\Theta),
\qquad
\chi'(r,\Theta)=\chi'_{\text{out}}(r,\Theta)=\frac{\omega}{\sqrt{A(r,\Theta)/B(r,\Theta)}}$,
and
$H^{(+)}_{\uparrow}(x,\Theta)=\mathbbm{i}\,Y^{(+)}_{\uparrow}(x,\Theta),
\qquad
\chi'(r,\Theta)=\chi'_{\text{in}}(r,\Theta)=-\frac{\omega}{\sqrt{A(r,\Theta)/B(r,\Theta)}}$.
The two branches are naturally interpreted as describing outward–moving and inward--moving modes, respectively, through the horizon \cite{vanzo2011tunnelling}.

Within this semiclassical picture, the tunneling probability associated with fermionic emission is determined by the imaginary part of the radial action. It can be written as
$\Gamma_{\psi}(\Theta,\omega)\sim
\exp\!\left[-2\,\text{Im}\!\left(\chi_{\text{out}}(r,\Theta)-\chi_{\text{in}}(r,\Theta)\right)\right].$
Using the symmetry between the two branches, the radial actions satisfy
$\chi_{\text{out}}(r,\Theta)=-\chi_{\text{in}}(r,\Theta)
=\int \mathrm{d}r\,\frac{\omega}{\sqrt{A(r,\Theta)/B(r,\Theta)}}$,
which completes the derivation of the tunneling rate in the present setup.

A subtle but relevant assumption enters at this stage of the analysis. The deformation parameter $\Theta$ is treated as a small quantity, a hypothesis that will later be corroborated by bounds obtained from solar system tests. Under this premise, the metric functions vary smoothly near the horizon and can be approximated by their first--order expansions around $r=r_h$.

This local linear behavior implies that the horizon contribution to the radial action is governed by a simple singular structure. As a result, the integrals that determine the imaginary part of the action acquire a first–order pole at $r=r_h$. Such a feature permits a straightforward evaluation using standard complex analysis prescriptions, leading directly to the expression written below
\ie
A{(r,\Theta)} \frac{1}{B{(r,\Theta)}} \approx \, A{(r,\Theta)\prime} \Bigg(\frac{1}{B{(r,\Theta) }}\Bigg)^{\prime}(r - r_{h})^{2} + ... \, .
\fe

By implementing Feynman’s prescription to regulate the pole at the horizon, the radial integral can be evaluated explicitly, leading to the result stated below:
\ie
2\mbox{Im}\;\left[  \chi_{ \text{out}}{(r,\Theta)} -  \chi_{ \text{in}} {(r,\Theta)}\right] =\mbox{Im}\int \mathrm{d} r \,\frac{4\omega}{\sqrt{\frac{A{(r,\Theta)}}{B{(r,\Theta)}}}}=\frac{2\pi\omega}{\kappa(r,\Theta)},
\fe
with $\kappa$ being given by
\ie
\kappa(r,\Theta) \Big|_{r=r_{h}} \approx {\, \, \frac{\Theta ^2 M \left(162 M^3-175 M^2 r_{h}+59 M r_{h}^2-6 r_{h}^3\right)}{16 r_{h}^7}+\frac{M}{r_{h}^2} = \frac{1}{4 M}.}
\fe

For fermionic modes propagating on the black hole background, the tunneling rate fixes the occupation factor of the emitted particles. This rate takes the Boltzmann-type form
$\Gamma_{\psi}(\omega,\Theta)\sim \exp\!\left(-\dfrac{2\pi\,\omega}{\kappa(r,\Theta)}\right)$. Replacing $\kappa(r,\Theta)$ by its value evaluated at the horizon position $r=r_h$ leads to the expression given below
\ie
\begin{split}
\label{fermmiiiipartii}
n_{\psi}(\omega,\Theta) & = {\frac{{\Gamma}_{\psi}(\omega, \Theta)}{1 + {\Gamma}_{\psi}(\omega,\Theta)}} = \, \, {\frac{1}{e^{8 \pi  M \omega }+1}} .
\end{split}
\fe

{The Klein--Gordon and Dirac calculations therefore lead to the same near-horizon exponential factor. The corresponding occupation numbers differ only through the statistical sign in the denominator: the minus sign applies to bosons, whereas the plus sign applies to fermions. The present result also reproduces the standard thermal contribution obtained through the radial-tunneling approach. The additional nonthermal terms found in the preceding subsection are associated with the implementation of energy conservation and backreaction through the replacement $M\rightarrow M-\omega'$, which is not included in the present fixed-background analysis.}


{\subsection{Particle production for fermions: the tunneling method} }

{

The leading-order WKB system obtained from the Dirac equation implies that the fermionic action satisfies the same Hamilton--Jacobi equation as the scalar action,
\begin{equation}
g^{\mu\nu}\partial_{\mu}I_{\psi}\partial_{\nu}I_{\psi} +m^{2}=0. \label{eq:fermionic-hj}
\end{equation}
Near the event horizon, the terms involving the fermion mass and angular momentum remain finite, whereas the radial contribution develops the pole responsible for the imaginary part of the action. Consequently, the outgoing fermionic mode follows the same radial characteristic employed in the bosonic tunneling calculation. The opposite spin polarization leads to the same radial equation and does not need to be considered separately.

To incorporate energy conservation and backreaction, we again perform the replacement
$M\longrightarrow \mu=M-\omega'$,
where $\omega'$ is the running energy of the emitted particle. The corresponding instantaneous horizon is located at
$r_{h}(\omega')=2\mu=2(M-\omega')$.
Using the outgoing radial velocity already obtained in the bosonic sector, the imaginary part of the fermionic action can be written as
\begin{equation}
\operatorname{Im}I_{\psi} = -\operatorname{Im} \int_{0}^{\omega}\mathrm{d}\omega' \int_{r_{i}}^{r_{f}} \frac{\mathrm{d}r} {\dot r_{\mathrm{out}}(r,\mu,\Theta)}. \label{eq:fermion-im-action}
\end{equation}

The perturbative expansion may be performed directly at the level of the radial integrand. For this purpose, let
$\dot r_{\mathrm{out}}(r,\mu,\Theta) = v_{0}(r,\mu) +\Theta^{2}v_{2}(r,\mu) +\mathcal{O}(\Theta^{4})$,
so that
\begin{equation}
\frac{1}{\dot r_{\mathrm{out}}} = \frac{1}{v_{0}} -\Theta^{2}\frac{v_{2}}{v_{0}^{2}} +\mathcal{O}(\Theta^{4}). \label{eq:fermion-integrand-expansion}
\end{equation}
Close to $r=2\mu$, the two contributions can be expressed as
\begin{equation}
v_{0}(r,\mu) = a_{0}(\mu)(r-2\mu) +\mathcal{O}\!\left((r-2\mu)^{2}\right),
\end{equation}
and
\begin{equation}
v_{2}(r,\mu) = a_{2}(\mu)(r-2\mu) +\mathcal{O}\!\left((r-2\mu)^{2}\right).
\end{equation}
It follows that the residue is determined by
\begin{equation}
\operatorname*{Res}_{r=2\mu} \left(\frac{1}{\dot r_{\mathrm{out}}}\right) = \frac{1}{a_{0}(\mu)} - \Theta^{2} \frac{a_{2}(\mu)}{a_{0}^{2}(\mu)} +\mathcal{O}(\Theta^{4}). \label{eq:general-fermion-residue}
\end{equation}

To evaluate these coefficients consistently, it is convenient to introduce
\begin{equation}
F(r,\Theta)=\frac{1}{B(r,\Theta)}.
\end{equation}
The complete outgoing radial velocity behaves near the instantaneous horizon as
\begin{equation}
\dot r_{\mathrm{out}} \simeq \frac{1}{2} \sqrt{A(r,\Theta)F(r,\Theta)} \simeq \kappa(\mu,\Theta)(r-2\mu), \label{eq:fermionic-radial-near-horizon}
\end{equation}
where
\begin{equation}
\kappa(\mu,\Theta) = \frac{1}{2} \sqrt{ A'(2\mu,\Theta)F'(2\mu,\Theta) }. \label{eq:instantaneous-kappa}
\end{equation}
After implementing $M\rightarrow\mu$, the derivatives evaluated at the instantaneous horizon are
\begin{equation}
A'(2\mu,\Theta) = \frac{1}{2\mu} + \frac{\Theta^{2}}{256\mu^{3}},
\end{equation}
and
\begin{equation}
F'(2\mu,\Theta) = \frac{1}{2\mu} - \frac{\Theta^{2}}{256\mu^{3}}.
\end{equation}
The contributions of second order in $\Theta$ cancel in their product, yielding
\begin{equation}
\kappa(\mu,\Theta) = \frac{1}{4\mu} +\mathcal{O}(\Theta^{4}).
\label{eq:instantaneous-kappa-result}
\end{equation}
Accordingly,
\begin{equation}
\dot r_{\mathrm{out}} \simeq \frac{r-2\mu}{4\mu} +\mathcal{O}(\Theta^{4}),
\end{equation}
and the residue of the radial integrand becomes
\begin{equation}
\operatorname*{Res}_{r=2\mu} \left(\frac{1}{\dot r_{\mathrm{out}}}\right) = 4\mu+\mathcal{O}(\Theta^{4}).
\label{eq:fermion-residue}
\end{equation}

Applying the Feynman prescription to the pole in Eq.~\eqref{eq:fermion-im-action}, the radial integration gives
\begin{align}
\operatorname{Im}I_{\psi} &= \pi\int_{0}^{\omega} \left[ 4(M-\omega') +\mathcal{O}(\Theta^{4}) \right]\mathrm{d}\omega'  =2\pi\omega(2M-\omega) +\mathcal{O}(\Theta^{4}).
\label{eq:fermion-final-action}
\end{align}
The tunneling probability for fermionic emission is therefore
\begin{equation}
\Gamma_{\psi}(\omega,\Theta) \sim \exp\left[ -4\pi\omega(2M-\omega) +\mathcal{O}(\Theta^{4}) \right].
\label{eq:fermion-tunneling-rate}
\end{equation}
The corresponding occupation number follows the Fermi--Dirac prescription,
\begin{equation}
n_{\psi}(\omega,\Theta) = \frac{\Gamma_{\psi}(\omega,\Theta)} {1+\Gamma_{\psi}(\omega,\Theta)} = \frac{1}{ \exp\left[ 4\pi\omega(2M-\omega) +\mathcal{O}(\Theta^{4}) \right]+1 }.
\label{eq:fermion-occupation-backreaction}
\end{equation}

The exponential factor is identical to the one governing bosonic tunneling, since it is fixed by the same horizon pole and the same energy-conservation prescription. The distinction between the two sectors arises solely from the statistical sign in the occupation number. Moreover, the term quadratic in $\omega$ represents the nonthermal correction induced by backreaction and disappears in the fixed-background limit $\omega\ll M$, for which the standard thermal factor $8\pi M\omega$ is recovered.

It is important that the perturbative expansion be applied to the complete outgoing radial velocity before its Laurent expansion around $r=2\mu$. Expanding separately quantities that become singular at the horizon may produce an artificial correction of order $\Theta^{2}$ in the residue. For the present geometry, the cancellation in Eq.~\eqref{eq:instantaneous-kappa-result} shows that no such correction survives in the horizon tunneling exponent at the perturbative order considered.

}

\section{Black hole evaporation and radiative emission }

This section examines the evaporation process of the black hole at a qualitative level. The discussion relies on the use of the \textit{Stefan--Boltzmann} relation as a practical tool to estimate the radiative energy loss associated with thermal emission, following the standard approach adopted in previous studies \cite{ong2018effective,12aa2025particle}
\ie
\label{sflaw}
\frac{\mathrm{d}M}{\mathrm{d}t}  =  - \alpha a {\Tilde{\sigma}} T^{4}.
\fe

Here, $\widetilde{\sigma}$ is interpreted as the absorption cross section of the  black hole under consideration. For the massless scalar field  considered in the bosonic sector, the radiated power can be obtained directly  from the Klein--Gordon equation and written in terms of the corresponding  greybody factors as
\begin{equation}
P_{\Phi}(\Theta) \equiv -\frac{\mathrm{d}M}{\mathrm{d}t} = \sum_{\lambda} \int_{0}^{\infty} \frac{\mathrm{d}\omega}{2\pi}\, \frac{\omega\,\Gamma_{\lambda}(\omega,\Theta)} {\exp\!\left(\omega/T_{\mathrm{H}}\right)-1},
\label{scalar_hawking_power}
\end{equation}
where $\lambda$ collectively labels the scattering channels and $\Gamma_{\lambda}(\omega,\Theta)$ denotes the corresponding transmission probability. These quantities are determined by solving the radial scattering problem obtained from the Klein--Gordon equation, imposing a purely ingoing solution at the horizon and a superposition of incident and reflected waves at spatial infinity. In other words, the greybody factors are not independent phenomenological constants, but are fixed by the propagation of the scalar field on the background geometry \cite{page1976particle,hiscock1990evolution}.

Nevertheless, the connection with the geometric--optics approximation can also be established directly from the Klein--Gordon equation. Introducing the eikonal form $\Phi\simeq\mathcal{A}\exp(iS/\epsilon)$ and retaining the leading contribution as $\epsilon\rightarrow0$ gives
\begin{equation}
g^{\mu\nu}\partial_{\mu}S\,\partial_{\nu}S=0.
\label{scalar_eikonal_equation}
\end{equation}
In this manner, the high--frequency scalar modes propagate along null geodesics. In this regime, the transmission probability (or greybody factors) approaches unity \cite{liang2025einstein} for modes whose impact parameter lies below the critical value $b_{\mathrm{c}}(\Theta)$ and vanishes for modes outside the capture region. The sum over the transmitted modes then reduces to the geometric capture cross section,
\begin{equation}
\widetilde{\sigma}_{\mathrm{lim}}(\Theta) = \pi b_{\mathrm{c}}^{\,2}(\Theta) = \pi R_{\mathrm{sh}}^{\,2}(\Theta).
\label{geometric_cross_section}
\end{equation}
{
For an axisymmetric geometry, the geometric capture cross section for a specified incidence direction is the area enclosed by the corresponding capture boundary on the impact parameter plane. It cannot, in general, be obtained from the equatorial critical impact parameter alone. Moreover, the total Hawking luminosity involves the absorption cross section averaged over incidence directions. In the estimates below, we introduce the additional approximation of replacing this averaged capture area by $\widetilde{\sigma}_{\mathrm{eff}}\equiv\pi b_{\mathrm{eq}}^{2}$. This prescription provides a simple equatorial estimate for the evaporation and emission calculations, but it does not constitute a determination of the full axisymmetric capture cross section. 
}

For the configuration adopted in the present analysis, the shadow radius expanded up to second order in $\Theta$ is
\begin{equation}
R_{\mathrm{sh}}(\Theta) \approx \, 3\sqrt{3}\,M + \frac{5\Theta^{2}}{288\sqrt{3}\,M} + \mathcal{O}\!\left(\Theta^{4}\right).
\label{shadow_radius_expansion}
\end{equation}

{
The geometrical meaning of this quantity deserves clarification, since the spacetime is axisymmetric and the equatorial calculation does not determine the entire shadow. Let $D_{\mathrm{eq}}(r,\Theta)=g_{\phi\phi}(r,\pi/2,\Theta)$ and $A(r,\Theta)=-g_{tt}(r,\Theta)$. The reflection symmetry about $\theta=\pi/2$ allows null geodesics initially tangent to the equatorial plane to remain in that plane. For these trajectories, the conserved quantities are $E = A\dot{t}$ and $L_z = D_{\mathrm{eq}}\dot{\phi}$, and the radial equation becomes $AB\dot{r}^{2} = E^{2}-AL_z^{2}/D_{\mathrm{eq}}$. The unstable circular orbit is determined by $\left.\partial_r(D_{\mathrm{eq}}/A)\right|_{r = r_{\mathrm{eq}}} = 0$, with $b_{\mathrm{eq}}^{2} = (L_z/E)^{2} = D_{\mathrm{eq}}(r_{\mathrm{eq}},\Theta)/A(r_{\mathrm{eq}},\Theta)$. Expanding these relations through second order gives $r_{\mathrm{eq}}=3M+\Theta^{2}/(1296M)+\mathcal{O}(\Theta^{4})$ and $b_{\mathrm{eq}}^{2}=27M^{2}+5\Theta^{2}/48+\mathcal{O}(\Theta^{4})$, which reproduces Eq.~(\ref{shadow_radius_expansion}).

In addition, the distinction from a circular shadow can also be seen by considering meridional trajectories, for which $L_z=0$ and $\phi$ is constant. In this sector, $p_{\theta}=C\dot{\theta}$ is conserved, where $C(r,\Theta)=g_{\theta\theta}(r,\Theta)$, and the corresponding critical impact parameter follows from the stationary value of $C/A$. We obtain $r_{\mathrm{mer}}=3M-\Theta^{2}/(324M)+\mathcal{O}(\Theta^{4})$ and $b_{\mathrm{mer}}^{2}=27M^{2}+\mathcal{O}(\Theta^{4})$. For a static observer at infinity in the equatorial plane, $b_{\mathrm{eq}}$ and $b_{\mathrm{mer}}$ give the magnitudes of the horizontal and vertical intercepts of the shadow boundary, respectively. Their different second order corrections show that the shadow is not circular for this viewing configuration. These two intercepts do not determine the complete contour, whose calculation requires treating generic null trajectories and specifying the observer's inclination.
}

It follows explicitly that
\begin{align}
\widetilde{\sigma}_{\mathrm{lim}}(\Theta) & \approx \pi \left( 3\sqrt{3}\,M + \frac{5\Theta^{2}}{288\sqrt{3}\,M} \right)^{2}  = 27\pi M^{2} + \frac{5\pi\Theta^{2}}{48} + \mathcal{O}\!\left(\Theta^{4}\right).
\label{limiting_cross_section}
\end{align}

For one real massless scalar degree of freedom, the geometric--optics limit of
Eq.~\eqref{scalar_hawking_power} may equivalently be expressed as
\begin{equation}
P_{\Phi}^{\mathrm{GO}} = \widetilde{\sigma}_{\mathrm{lim}} \int_{0}^{\infty} \frac{\mathrm{d}\omega}{2\pi^{2}}\, \frac{\omega^{3}} {\exp\!\left(\omega/T_{\mathrm{H}}\right)-1} = \frac{\pi^{2}}{30}\, \widetilde{\sigma}_{\mathrm{lim}}T^{4},
\label{geometric_optics_power}
\end{equation}
where $T=1/(8\pi M)$. Therefore, when $\widetilde{\sigma}_{\mathrm{lim}}$ denotes an absorption cross section, the coefficient multiplying $\widetilde{\sigma}_{\mathrm{lim}}T^{4}$ is fixed from first principles as $\pi^{2}/30$ for a single massless scalar degree of freedom. In addition, contributions from additional particle species (the most general case) must instead be included by summing their respective spin--dependent Hawking spectra and greybody factors.

{
Here, $T_H=1/(8\pi M)$, and the calculation refers to one real massless scalar degree of freedom. To make the normalization explicit, we use the identity $\int_{0}^{\infty}\omega^{3}(e^{\omega/T_H}-1)^{-1}\,\mathrm{d}\omega = \pi^{4}T_H^{4}/15$, which gives $P_{\Phi}^{\mathrm{GO}} =(\pi^{2}/30)\widetilde{\sigma}_{\mathrm{lim}}T_H^{4}$ when combined with the factor $1/(2\pi^{2})$ in Eq.~(\ref{geometric_optics_power}). In this manner, when $\widetilde{\sigma}_{\mathrm{lim}}$ represents the absorption cross section, the notation introduced in Eq.~(\ref{sflaw}) requires $a\alpha=\pi^{2}/30$ for the scalar field considered here. This coefficient is fixed by the spectral integral and is used naturally in the subsequent evaporation calculation. 
}

The existence of thermal emission from black holes was established in the seminal work of Ref.~\cite{hawking1975particle}, where the radiation spectrum was shown to be characterized by a well--defined temperature. While equivalent derivations based on Bogoliubov coefficients are available in the literature, they are not pursued here. Within the framework developed above, the relation in Eq.~(\ref{sflaw}) can therefore be rewritten as follows:
\ie
\begin{split}
\label{dmmm}
\frac{\mathrm{d}M}{\mathrm{d}t} = & - {\frac{\pi^{2}}{30}} \frac{ \left(\frac{5 \pi  \Theta ^2}{48}+27 \pi  M^2\right)}{4096 \pi ^4 M^4} \, ,
\end{split}
\fe
{Retaining terms through second order in $\Theta$ and using the normalization $a\alpha=\pi^{2}/30$ for one real massless scalar degree of freedom, we integrate the mass loss equation as follows:}
\ie
\begin{split}
\label{dmmeee}
 \int_{0}^{t_{\text{evap}}} \mathrm{d}\tau  & =  \int_{M_{i}}^{M_{f}} 
\left[ - {\frac{\pi^{2}}{30}} \frac{ \left(\frac{5 \pi  \Theta ^2}{48}+27 \pi  M^2\right)}{4096 \pi ^4 M^4}  \right]^{-1} \mathrm{d}M.
\end{split}
\fe
The quantity $t_{\text{evap}}$ represents the complete lifetime of the black hole. Within the approximations adopted here, this timescale is expressed as:
\ie
\begin{split}
t_{\text{evap}} \approx & \, \, \frac{{12800} {\pi} \Theta ^2 (M_{f}-M_{i})}{{729}}- {\frac{40960}{27}} \left( {\pi} \left(M_{f}^3-M_{i}^3\right)\right),
\end{split}
\fe
where contributions beyond second order in $\Theta$ have also been discarded.

The evaporation timescale $t_{\text{evap}}$ follows from a fully analytic computation. {Within the perturbative semiclassical approximation adopted here, neither the Hawking temperature nor the mass loss rate vanishes at a finite positive mass. Therefore, no finite mass evaporation endpoint is found at the perturbative order considered. Nevertheless, this result does not constitute a definitive proof of the absence of a stable remnant, since the expansion is valid only in the regime $\Theta^{2}/M^{2}\ll1$ and cannot reliably describe the final stage in which the black hole mass becomes comparable to the NC scale. Accordingly, the limit $M_f\to0$ is adopted below as a formal extrapolation of the semiclassical mass loss law, instead of a definitive (final) statement concerning the exact endpoint of the evaporation process. Under this assumption, we obtain:}
\ie
t_{\text{evap}} = \, \, \frac{{40960 \pi}M_{i}^3}{ {27} } -\frac{{12800 \pi} \Theta ^2 M_{i}}{{729}}.
\fe

\begin{figure}
    \centering
     \includegraphics[scale=0.51]{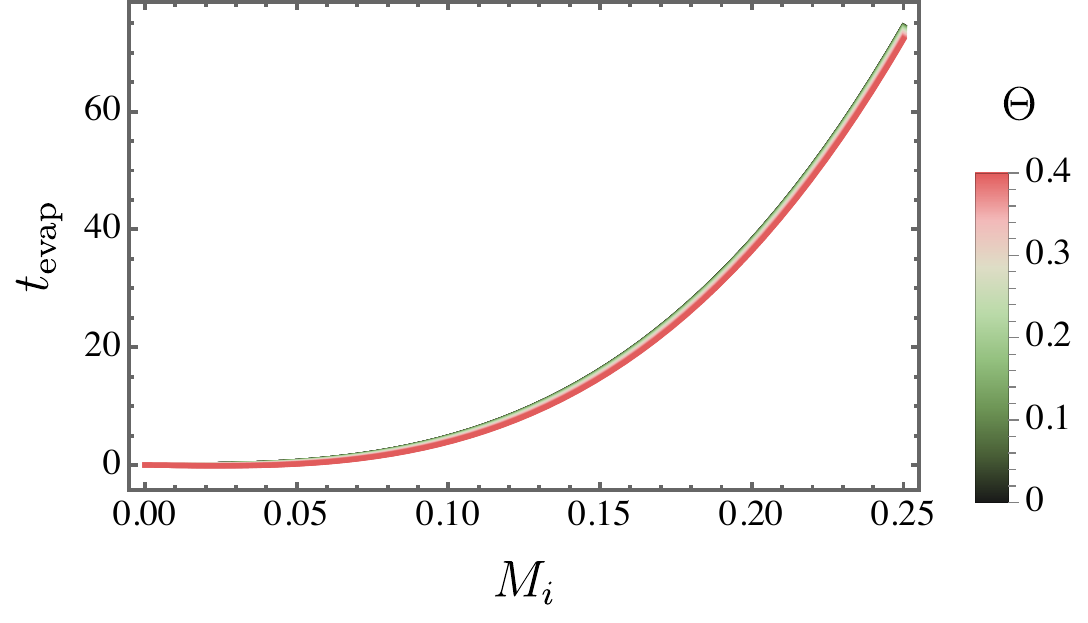}
    \caption{Total evaporation time $t_{\text{evap}}$ as a function of the initial mass $M_i$ for several values of the {NC} parameter $\Theta$.}
    \label{evapp}
\end{figure}

Figure~\ref{evapp} illustrates the evaporation behavior and shows that increasing the {NC} parameter leads to a shorter black--hole lifetime. This trend is consistent with results reported for other {NC} black hole models \cite{Heidari:2025oop,AraujoFilho:2025jcu}. The same tendency is confirmed below through the analysis of the corresponding emission rate.

Near the event horizon, quantum phenomena become dominant and allow vacuum fluctuations to generate short–lived particle pairs. In certain instances, one member of the pair acquires sufficient energy to penetrate the effective barrier and propagate to infinity through a tunneling process. The escape of this positive energy mode reduces the mass of the black hole and drives its evaporation, a mechanism identified with Hawking radiation. From the perspective of a distant observer, this mass loss is encoded in the high--energy absorption properties of the spacetime. In this limit, the absorption cross section approaches a constant value, conventionally denoted by ${\tilde{\sigma}}_{lim}$. As shown in Refs.~\cite{decanini2011universality,papnoi2022rotating}, this asymptotic behavior controls the form of the spectral energy flux. {To maintain the same normalization used in the evaporation calculation, we consider the spectral energy emission rate for one real massless scalar degree of freedom. Within the geometric--optics approximation, this quantity is given by}
\ie
\label{emission}
	\frac{\mathrm{d}^{2}E}{\mathrm{d}\omega\,\mathrm{d}t} = \frac{\widetilde{\sigma}_{lim}}{{2\pi^{2}}}\frac{\omega^{3}}{e^{\omega/T_H}-1}.
\fe
{ where $\omega$ denotes the angular frequency, or equivalently the energy of the emitted scalar quantum in the natural units adopted throughout this work. Substituting $\widetilde{\sigma}_{\mathrm{lim}}=27\pi M^{2}+5\pi\Theta^{2}/48+\mathcal{O}(\Theta^{4})$ and $T_H=1/(8\pi M)+\mathcal{O}(\Theta^{4})$, we obtain }
\ie
\frac{\mathrm{d}^{2}E}{\mathrm{d}\omega \mathrm{d} t} = \frac{ \omega ^3 \left(5 \Theta ^2+1296 M^2\right)}{{96 \pi } \left(e^{8 \pi  M \omega }-1\right)}.
\fe

\begin{figure}
    \centering
     \includegraphics[scale=0.51]{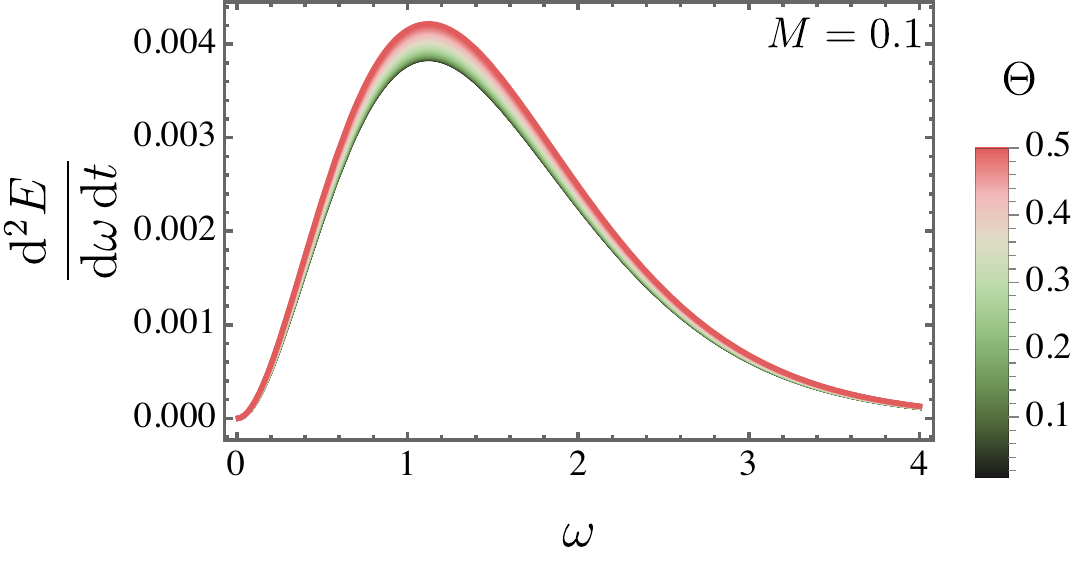}
    \caption{Emission rate as a function of the frequency $\omega$ for different values of the {NC} parameter $\Theta$.}
    \label{emmii}
\end{figure}

{
Figure~\ref{emmii} illustrates the spectral energy emission rate obtained with this normalization. At fixed $M$, the NC contribution enters through the effective absorption cross section, while the Hawking temperature remains unchanged through second order in $\Theta$. Within the geometric-optics approximation, the spectrum differs from its Schwarzschild counterpart by the factor $1+5\Theta^{2}/(1296M^{2})+\mathcal{O}(\Theta^{4})$. The correction increases the amplitude without changing the frequency of the spectral maximum at this order. The same relative enhancement appears in the integrated power, with the reduction of the evaporation time obtained above.
}

The results were obtained in natural units, $G=c=\hbar=k_{\mathrm B}=1$. The magnitude of the NC correction is controlled by the ratio $\Theta/M$, and the perturbative character of the metric requires $\Theta^{2}/M^{2}\ll1$. Therefore, NC effects are strongly suppressed for macroscopic black holes. In this manner, we have chosen relatively large values of $\Theta$ solely to improve the visualization of the plots and to make its influence on our results more clearly distinguishable. Primordial or microscopic black holes provide a more natural scenario for investigating such effects, since their characteristic size may become closer to the NC scale during evaporation. Nevertheless, when $M$ becomes comparable to $\Theta$, the perturbative approximation is no longer valid. Consequently, the present results cannot describe the final stages of evaporation or determine a possible endpoint, but only indicate the behavior of the NC corrections within the controlled semiclassical regime.

Our model is related to the general framework of Ref.~\cite{szabo2003quantum} through the Moyal deformation and the Seiberg--Witten map. In both cases, noncommutativity introduces a fundamental short distance scale and becomes more relevant when the characteristic physical scale approaches the NC scale. However, Ref.~\cite{szabo2003quantum} studies quantum field theories directly formulated on NC spaces, including nonplanar contributions and UV/IR mixing, whereas here the NC corrections are incorporated perturbatively into the gravitational background and Hawking radiation is treated semiclassically. Quantum loop effects such as UV/IR mixing are therefore beyond the scope of the present analysis.

\section{Bounds }

We can constrain the non--commutativity parameter by analyzing the classical tests of general relativity and estimating the magnitude of $\Theta$ that do not contradicts the experimental data. In order to do this, we consider the trajectory of Mercury as a test particle that follows a time-like geodesic and of light ray emitted by a distant source and bended by the Sun, and that reflected by the Cassini probe when grazing the surface of the Sun.

In the equatorial plane ($\theta=\pi/2$), the Lagrangian of test particles is given by

\begin{equation}\label{eq:lagr1}
{A}(r){{\dot t}^2} - B^{(\Theta)}{{\dot r}^2} - D^{(\Theta)}{{\dot \varphi }^2} = \eta=-2{\cal L}\, ,
\end{equation}
where ${\cal L}$ is the Lagrangian. When $\eta=1$, we have time-like trajectories, and when $\eta=0$ we have light-like trajectories. The conserved quantities, named energy $E$ and angular momentum $L$ are 

\begin{equation}\label{constant}
E = A^{(\Theta)}\dot t \quad\mathrm{and}\quad L = D^{(\Theta)}\dot \varphi.
\end{equation}

By working withe these conserved quantities, we can manipulate equation \eqref{eq:lagr1} to find a second order differential equation for the variable $u=L^2/(Mr)$ as

\begin{align}\label{eq:u}
     u''(\varphi)= \, \eta -u+3 \frac{M^2}{L^2} u^2 \left(1+\frac{\Theta ^2}{2L^2} \left(E^2-\eta \right)\right)\, .
\end{align}

We can see that the {NC} parameter couples with the energy of the test particle, which is a property found in several quantum gravity scenarios that lead to interesting phenomenological opportunities \cite{Addazi:2021xuf,AlvesBatista:2023wqm}.


\subsection{Perihelion precession of Mercury }

From \eqref{eq:u}, we verify that if we fix $\eta=1$, we are effectively redefining the mass of the central object that is responsible for the trajectory of the planet. This means that Mercury behaves as it is effectively surrounding a massive object of mass square
\begin{equation}
    \widetilde{M}^2=M^2\left(1+\frac{\Theta ^2}{2L^2} \left(E^2-\eta \right)\right).
\end{equation}

To constrain the parameter $\Theta$, we analyze the dimensionless perturbative term $\frac{\Theta^2}{2L^2}(E^2 - \eta)$.  
In a Keplerian orbit, the specific angular momentum $L$ relates to the semi-major axis $a$ and eccentricity $e$ via $L^2 = M a (1 - e^2)$, and the specific orbital energy per unit mass is $E = -M/(2a)$ \cite{Goldstein:2002}.

{In the following we will use natural units, i.e., using units in terms of Planckian quantities, like the Planck mass $M_{\rm Pl}$, energy $E_{\rm Pl}$ and length $\ell_{\rm Pl}$.} Adopting natural units, we assign the Sun’s mass in term of $M_{\rm Pl}$ as $M_\odot = 9.138 \times 10^{37}$.  
For Mercury, we use $a = 3.583 \times 10^{45}$ in terms of the Planck lenght $\ell_{\rm}$ and $e = 0.2056$.  
From these, we compute $L = 5.600 \times 10^{41}$, which justifies a perturbative treatment because $M^2/L^2$ is negligibly small.  
We also find $E^2 = 1.627 \times 10^{-16}$, confirming that corrections depending on energy are insignificant.

The general relativistic prediction for Mercury’s perihelion advance is $\Delta\Phi_{\text{GR}} = 42.9814''$ per century.  
The observed value, $\Delta\Phi_{\text{Exp}} = (42.9794 \pm 0.0030)''$ per century \cite{Casana:2017jkc,Yang:2023wtu}, agrees exceptionally well.  
By comparing the experimental uncertainty with our theoretical expression, we derive the allowed range 
\[
\Theta^2 \le 1.862 \times 10^{10} \; \text{m}^2.\]

{The {NC} correction is suppressed by the angular momentum of Mercury. This places a challenge for constraining quantum gravity parameters using Solar System tests. If the produced spacetime presented a correction suppressed by the mass of the planet, the bound would have been better. It is expected that a quantity with dimension of area should suppress terms that depend on $\Theta^2$. Maybe other choices of Moyal twist could produce a metric that could probe the deformation scale with better sensitivity.}


\subsection{Deflection of light }

For the deflection of light, we consider the case $\eta=0$, but we work with the function $u=1/r$. In this case, the dimensionless correction is given by $\Theta^2/(2b^2)$, where $b=L/E$ is the impact parameter of the light ray.

To derive constraints from light deflection, we consider a photon grazing the solar limb, for which the impact parameter is well approximated by the solar radius: $b \simeq R_{\odot} = 4.305 \times 10^{43}$ in natural units. We adopt the solar mass $M_\odot = 9.138 \times 10^{37}$.

In General Relativity, the bending angle for such a configuration is $\delta_{\mathrm{GR}} = 4M_\odot/b = 1.7516687''$. Experimental results are commonly parametrized as $\delta_{\mathrm{exp}} = \frac{1}{2}(1 + \gamma) \times 1.7516687''$, where the parameter $\gamma$ is measured to be $0.99992 \pm 0.00012$ \cite{dsasdas}.

The effect of the parameter $\Theta$ enters through a multiplicative factor $1 - \Theta^2/(4b^2)$. Matching this factor to the observed value $(1+\gamma)/2$ results in the bound
\[
\Theta^2 \; \leq \; 1.936 \times 10^{13} \; \mathrm{m}^2 .
\]

The {NC} effect is suppressed by the radius of the Sun, which places a severe difficulty for probing the deformation scale. { In addition, the light deflection has also been investigated in Kalb--Ramond--ModMax black hole geometries, where the additional fields modify the optical properties of the spacetime~\cite{Sucu:2026nkw}.
}


\subsection{Time delay of light }

To compute the Shapiro delay or time delay of light \cite{Shapiro:1964uw}, we need also to consider light-like geodesics, but parametrized by the coordinate time $t$. This can be achieved by working with equation \eqref{eq:lagr1} as

\begin{equation}
  (\dot{r})^2=\left(  \frac{\mathrm{d}r}{\mathrm{d}t}\right)^2=\frac{A^{(\Theta)}D^{(\Theta)}-\frac{\ell^2}{E^2}{(A^{(\Theta)})}^2}{B^{(\Theta)}D^{(\Theta)}}\, .
\end{equation}

We consider that a light ray emitted from the Earth passes close to the surface of the Sun, reaches an obstacle and returns to Earth. The coordinate of minimum distance to the Sun is a turning point characterized by the condition $\dot{r}=0$. Following Ref.~\cite{Wang:2024fiz}, we express the energy and angular momentum in terms of functions of this minimum distance $r_{\text{min}}$ as $L^2/E^2 = D^{(r,\Theta)}(r_{\text{min}})/A^{(r,\Theta)}(r_{\text{min}})$. For this reason, the time of propagation of the light ray can be written as
\begin{equation}\label{eq:shapiro_main}
    \mathrm{d} t=\pm \frac{1}{A^{(\Theta)}}\frac{1}{\sqrt{\frac{1}{A^{(\Theta)}B^{(\Theta)}}-\frac{D^{(\Theta)}(r_{\text{min}})/A^{(\Theta)}(r_{\text{min}})}{B^{(\Theta)}D^{(\Theta)}}}}\, .
\end{equation}

The integration of this expression leads to the approximate expression

\begin{align}
    t=\sqrt{r^2-r_{\text{min}}^2}+M\left(\sqrt{\frac{r-r_{\text{min}}}{r+r_{\text{min}}}}+2\ln\left(\frac{r+\sqrt{r^2-r_{\text{min}}^2}}{r_{\text{min}}}\right)\right)\\
    -\Theta ^2\frac{M}{8r_{\text{min}}r}\sqrt{\frac{r-r_{\text{min}}}{r+r_{\text{min}}}}\, .\nonumber
\end{align}

When $r \gg r_{\text{min}}$, we find

\begin{equation}\label{eq:sh_t_nc}
    t(r)=r+M+2M\ln\left(\frac{2r}{r_{\text{min}}}\right)-\Theta^2\frac{M}{8r_{\text{min}}r}\, .
\end{equation}

If $t(r_{E,R})$ is the time taken for a signal to travel from the emitter/Sun to the Sun/receiver. Then after the time taken for the full-round consisting of emission of the ray from the Earth, receiving at a spacecraft, and back to the Earth is

\begin{equation}
T_{\Theta}=2(r_E+r_R)+4M\left[1+\ln\left(\frac{4r_Rr_E}{r_{\text{min}}^2}\right)-\frac{\Theta^2}{8 r_{\text{min}}}\left(\frac{1}{r_E}+\frac{1}{r_R}\right)\right]=T_{\text{flat}}+\delta T\, .
\end{equation}

In the parametrized post-Newtonian formalism, the contribution to this effect is
\begin{equation}
    \delta T = 4M\left(1+\frac{1+\gamma}{2}\ln \left(\frac{4r_Rr_E}{r_{\text{min}}^2}\right)\right)\, .
\end{equation}

Data from the Cassini mission \cite{Bertotti:2003rm,Will:2014kxa} constrain the parameter $\gamma$ as $|\gamma - 1| < 2.3 \times 10^{-5}$. Let us assume the distance $r_E$ as the one between the Earth and the Sun, i.e., as one astronomical unit $r_E=1\, \text{AU}$. Also, when the measurement that set this bound on $\gamma$ was performed, the Cassini probe was located at $r_R = 8.46\, \text{AU}$ from the Sun, and the minimal radial distance was $r_{\text{min}}=1.6\, R_{\odot}$, where $R_{\odot}$ is the solar radius. Endowed with these tools and using the uncertainty in the determination of $\gamma$, we are led to the following constrain on the {NC} parameter
\begin{equation*}
    \Theta^2 \leq 1.565 \times 10^{18}\, \text{m}^2 \, .
\end{equation*}

{The term $r_{\rm min}/(r_E^{-1}+r_R^{-1})$ is larger than the area of the Sun, which is the reason why the constraint derived from this observable is weaker than the one from the deflection of light.}

A presentation of the resulting parameter bounds is given in Table~\ref{tab:constr}. {The bounds found from this choice of Moyal twist are weaker or at the same orther than other scenarios, for instance, from \cite{AraujoFilho:2025jcu}. The reason lies in the quantities with area dimension that suppress the {NC} corrections, which are larger than in previous approaches. Usually, the deflection of light indeed presents a dimensionless correction of the order $\Theta^2/b^2$, which furnishes bounds of the order $\sim 10^{13}-10^{14}\, \rm m^2$. In our case, Mercury's perihelion precession's main correction is of the order $\Theta^2/L^2$. On the other hand, in \cite{AraujoFilho:2025jcu}, this correction is given by $\Theta^2/M^2$, which furnishes a bound of the order $\sim 10^3-10^4\, \rm m^2$. Our Shapiro delay has a correction of the form $\sim \Theta^2/(1{\rm AU} R_{\odot})$ which is smaller than $\sim \Theta^2/R_{\odot}^2$. Therefore, in order to improve the bounds, it is necessary to have an area dimension suppression factor, found from a combination of the Solar System scales that is small enough to instead amplify the {NC} effects.}

\begin{table}[h!]
\centering
\caption{Bounds for $\Theta^2$ derived from Solar System tests.}
\label{tab:constr}
\begin{tabular}{lc}
\hline\hline
\textbf{Solar System Test} & \textbf{Constraints $(\text{m}^2)$} \\
\hline
Mercury precession   & $\Theta^2\leq 1.862\times 10^{10}$ \\
Light deflection     & $\Theta^2\leq 1.936\times 10^{13}$ \\
Shapiro time delay   & $\Theta^2 \leq 1.565 \times 10^{18}$ \\
\hline\hline
\end{tabular}
\end{table}


\section{Conclusion }

In this work, the quantum properties of a Schwarzschild--like black hole constructed within {NC} gauge gravity were investigated by adopting the Moyal twist $\partial_t \wedge \partial_\theta$, which ensured a well--defined surface gravity and {Hawking temperature (and Hawking radiation)}. The corrected construction of {NC} black hole geometries was taken as the starting point based on \cite{Juric:2025kjl}, and the resulting spacetime was shown to preserve the classical horizon location and surface gravity. {Consequently, the Hawking temperature remained equal to its Schwarzschild value at the perturbative order considered.}

Quantum particle creation was analyzed {for both bosonic and fermionic fields by employing two complementary approaches: the near--horizon method in a fixed background and the tunneling formalism including energy conservation and backreaction. The Klein--Gordon and Dirac analyses led to the same near--horizon exponential factor, with the corresponding occupation numbers differing only through the Bose--Einstein and Fermi--Dirac statistical signs. Moreover, when backreaction was incorporated through the replacement $M\rightarrow M-\omega'$, both sectors exhibited the same nonthermal correction proportional to $\omega^{2}$. However, the contributions of order $\Theta^{2}$ canceled in the surface gravity and in the residue of the horizon pole. Therefore, no correction associated with the NC parameter survived in the bosonic or fermionic tunneling exponent at the perturbative order considered.}

The evaporation dynamics were subsequently examined through an effective \textit{Stefan--Boltzmann} approach combined with the high--energy absorption cross section. It was found that the {NC} corrections enhanced the radiative emission rate and reduced the total evaporation time, in agreement with the qualitative behavior observed in other {NC} black hole scenarios. {Since the Hawking temperature remained unchanged, these modifications originated from the correction to the critical impact parameter and, equivalently, to the effective absorption cross section. Within the perturbative semiclassical regime, neither the temperature nor the mass--loss rate vanished at a finite positive mass, and therefore no finite--mass endpoint was found at the order considered. Nevertheless, because the condition $\Theta^{2}/M^{2}\ll1$ ceases to be valid during the final stages of evaporation, the limit $M_f\rightarrow0$ was adopted only as a formal extrapolation of the semiclassical mass--loss law. }

Finally, bounds on the {NC} parameter $\Theta$ were inferred from classical Solar System tests, namely the perihelion precession of Mercury, light deflection, and the Shapiro time delay. The resulting constraints restricted the allowed magnitude of $\Theta^2$ and justified the perturbative regime adopted in the analysis. {Among the observables considered, the perihelion precession of Mercury provided the strongest constraint, $\Theta^{2}\leq1.862\times10^{10}\,{\rm m}^{2}$, followed by light deflection and the Shapiro time delay.}

{Since the noncommutativity parameter is generally expected to be of the order of the Planck area, $\sim10^{-70}\,{\rm m}^{2}$, the bounds obtained here illustrate the severe difficulty of directly constraining quantum--gravity effects through Solar System tests. More sensitive approaches should therefore be pursued. In particular, quantum--gravity corrections may become more relevant for primordial or microscopic black holes, whose characteristic scales may approach the NC scale during evaporation. Possible modifications during the late stages of evaporation could leave imprints on primordial black hole populations and provide observational connections between quantum gravity, astrophysics, and cosmology \cite{Profumo:2026suy}. However, this regime lies outside the validity of the perturbative approximation employed here and remains beyond the scope of the present work.}


\section*{Acknowledgments}
\hspace{0.5cm} A.A.A.F. is supported by Conselho Nacional de Desenvolvimento Cient\'{\i}fico e Tecnol\'{o}gico (CNPq) and Fundação de Apoio à Pesquisa do Estado da Paraíba (FAPESQ), project numbers 150223/2025-0 and 1951/2025. I. P. L. was partially supported by the National Council for Scientific and Technological Development - CNPq, grant 312547/2023-4. I. P. L. would like to acknowledge networking support by the COST Action BridgeQG (CA23130), the COST Action RQI (CA23115) and the COST Action FuSe (CA24101) supported by COST (European Cooperation in Science and Technology). A.R.Q. also thanks the Para\'iba State Research Foundation (FAPESQ) for financial support.  A.R.Q. also acknowledges the support by CNPq under grant 310533/2022-8.

\section*{Data Availability Statement}

Data Availability Statement: No Data associated with the manuscript

\bibliographystyle{ieeetr}
\bibliography{main}

\end{document}